\documentstyle[twocolumn,aps]{revtex}
\include{epsf}

\begin{document}
\draft
%\preprint{HEP/123-qed}
\title{Quantum Computation in a
One-Dimensional Crystal Lattice \\ with NMR Force Microscopy}
\author{T. D.  Ladd, J. R. Goldman, A. D\^ana, F. Yamaguchi, and Y. Yamamoto \cite{byline}}
\address{Quantum Entanglement Project, ICORP, JST \\ Edward L.
Ginzton Laboratory, Stanford University, Stanford, California
94305-4085}
\date{\today}
\maketitle
\begin{abstract}
A proposal for a scalable, solid-state implementation of a quantum
%60 characters up to q in quantum.
computer is presented.  Qubits are fluorine nuclear spins in a
solid crystal of fluorapatite [Ca$_5$F(PO$_4$)$_3$] with resonant
frequencies separated by a large field gradient. Quantum logic is
accomplished using nuclear-nuclear dipolar couplings with
decoupling and selective recoupling RF pulse sequences. Magnetic
resonance force microscopy is used for readout.
%
%This proposal
%takes advantage of many of the successful aspects of solution NMR
%quantum computation, including ensemble measurement and long
%$T_1$, but it allows for more qubits and the potential for
%initialization.
%
% %%%%%%%%%%%%%%%%%%%%%%%%%%%%%%%%%%%%%%%%%%%%%%%%%%%%%%%%%
% %                                                       %
% % The above sentence was removed in order to shorten    %
% % the abstract to < 600 characters, as requested by PRL.%
% %                                                       %
% %%%%%%%%%%%%%%%%%%%%%%%%%%%%%%%%%%%%%%%%%%%%%%%%%%%%%%%%%
%
As many as 300 qubits can be implemented in the realistic
laboratory extremes of $T=10$~mK and $B_0=20$~T with the existing
sensitivity of force microscopy.
\end{abstract}

\narrowtext

An increasing number of theoretical developments have recently
motivated the construction of a quantum computer with a large
number of qubits.  The potential for efficient simulation of
quantum systems \cite{lloyd96}, the discoveries of algorithms for
fast factorization \cite{shor94} and database searching
\cite{grover97}, and the development of quantum error correction
\cite{shor95} have helped to spur a large number of proposals for
experimental implementations of a quantum computer. However, any
physical implementation of a quantum computer must battle the fact
that well-isolated quantum systems are difficult to couple and to
measure, whereas the introduction of necessary couplings and
probes leads to the devastating effects of decoherence.

To date, the most successful experimental realization of a
multi-qubit quantum computer (or at least a simulation thereof)
has been in room-temperature solution NMR \cite{nmrqc}. Here, the
spin-states of molecular nuclei in a solution are well isolated,
as demonstrated by long thermal relaxation times ($T_1$) of many
seconds.  The nuclei of each molecule are weakly coupled by scalar
couplings. Measurement without substantial decoherence is made
possible by the large ensemble of ${\sim}10^{22}$ uncoupled,
identical molecules in the solution. There is much debate,
however, as to whether solution NMR currently does or ever will
exhibit the signatures of truly {\it quantum} information
processing \cite{schack1999}.  The presence of a large, mixed
ensemble renders quantum entanglement and wave-function collapse
of individual spins unmeasurable. Certain manipulations allow the
processing of an ``effective pure state" in which the ensemble
behaves nearly identically to a single quantum system
\cite{nmrqc,knill98}.   However, it has been shown that the
evolution of effective pure states of existing solution NMR
computers can be described without any entanglement, and either
more qubits or much larger nuclear polarizations will be needed
before entanglement is demonstrable \cite{braunstein99}. Even if
we neglect the debate of what is ``quantum" and what is not, the
principal limitation due to the small polarizations in solution
NMR quantum computation is that the usable signal decreases
exponentially with the number of qubits, leaving this method
unlikely to exceed the 10 qubit level without extensive
modification \cite{warren97}.

In this letter, we propose an implementation using solid-state
NMR. In a solid crystal there exists the potential for
polarization of the nuclear spins by simple cooling of the lattice
to extremely low temperatures.  The thermal relaxation time $T_1$
can be dramatically lengthened when moving from a solution to an
insulating, nonmagnetic crystal, and the nuclear-nuclear coupling
can be faster.  Hence, a solid crystal has great potential for
quantum computation in terms of timescales \cite{ladd2000} and
scalability.

\begin{figure}
\begin{center}
\epsfysize=2in %\epsffile{figure1epsAA.eps}
\epsffile{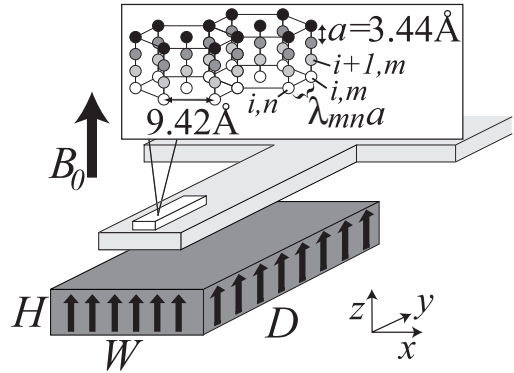}
\end{center}
\caption{ The fluorapatite crystal, of length $10~\mu$m and width
roughly $1~\mu$m, is mounted near the end of a cantilever. A
dysprosium micromagnet with dimensions $W=10~\mu$m, $H=4~\mu$m,
$D=400~\mu$m is aligned parallel to the crystal planes. A field
gradient of 1.4~T/$\mu$m is produced at the crystal, a distance of
2.07~$\mu$m above the magnet. The insert shows the $^{19}$F sites
of the crystal structure.  The different shades indicate different
resonant frequencies. \label{fig}}
\end{figure}

These advantages, however, come at a cost: several difficult and
related problems arise in implementing a solid-crystal NMR
computer.  This letter, although not the first proposal for
quantum computation in a crystal lattice \cite{yamaguchi99},
provides new approaches to these problems.  One problem is the
complicated network of dipolar couplings in a solid, which must be
suppressed to slow $T_2$ decoherence but selectively retained for
logic.  We address this problem with both ``selective averaging"
radio frequency (RF) pulse sequences and a well-chosen crystal
structure.   Another problem arises from the need to distinguish
and detect nuclei in the periodic lattice of a crystal; for this
we establish a very large, static, one-dimensional magnetic field
gradient with a microfabricated, high-magnetization ferromagnet
\cite{goldman2000}.  Only a small ensemble of nuclei may fit into
the area over which the gradient field is homogeneous, and
therefore a more sensitive means than standard inductive pick-up
is needed to measure the nuclear spin states. We propose using
magnetic resonance force microscopy (MRFM).

A schematic for the solid-crystal quantum computer we propose
appears in Fig. \ref{fig}. The quantum computer is an ensemble of
$N$ one-dimensional chains of $n$ spin-1/2 nuclei.  Due to the
field gradient, the resonant frequencies of the nuclear spins
within a chain are different.  Hence the secular component of the
dipolar Hamiltonian which couples the $i$th spin to the $j$th spin
within the $m$th chain is written \cite{slichter}
    \begin{eqnarray}\nonumber
    \hat{\cal H}_{ijmm}&=&\frac{\mu_0}{4\pi}\gamma^2\hbar^2
    \frac{1-3\cos^2\phi}{[|j-i|a]^3}\hat{I}_{im}^z \hat{I}_{jm}^z
    \\ &\equiv& \hbar\delta\omega_{ij}\hat{I}_{im}^z \hat{I}_{jm}^z,
    \label{ijcouple}\end{eqnarray}
where $\gamma$ is the gyromagnetic ratio ($2\pi\times 40$~MHz/T
for $^{19}$F), $a$ is the distance between spins in a chain, and
$\phi$ is the angle between the chain of spins and the large
applied magnetic field, which is taken to be in the $z$-direction.
Other terms of the dipolar Hamiltonian average to zero on a
timescale of 1/$\Delta\omega_{ij}$ or faster, where
$\Delta\omega_{ij}=\gamma a|\nabla B^z_{ij}|$ is the separation of
the $i$th and $j$th resonant frequencies caused by the field
gradient $\nabla B^z_{ij}$ between them.  The Hamiltonian of Eq.
(\ref{ijcouple}) may be ``switched off" by applying a periodic
succession of narrow band $\pi$ pulses at, for instance, the $i$th
resonant frequency \cite{slichter}. Simultaneous decoupling of
more than two qubits may be accomplished by timing the selective
$\pi$ pulses according to the entries of an appropriately sized
Hadamard matrix; a pair of qubits may be selectively recoupled in
order to implement two-bit gates \cite{leung99}.

Non-disturbing measurement is possible in this scheme because,
orthogonal to the chain direction, a nucleus of resonant frequency
$\omega_i$ has a large plane of copies with equal frequency. The
coupling between copies has a different form; the $n$th and $m$th
nucleus of identical resonant frequency $\omega_i$ are coupled by
the Hamiltonian \cite{slichter} $$
    \hat{\cal H}_{iimn}=\frac{1}{2}\frac{\mu_0}{4\pi}\gamma^2\hbar^2
    \frac{1-3\cos^2\theta_{mn}}{(a\lambda_{mn})^3}
    \left(3\hat{I}_{im}^z \hat{I}_{in}^z-{\hat {\bf I}}_{im}\cdot{\hat {\bf I}}_{in}\right),
$$
where $\lambda_{mn}$ is the distance between the two nuclei in
units of $a$, and $\theta_{mn}$ is the angle between the vector
which connects them and the direction of the applied field. This
coupling between copies must be ``switched off" to prevent
interference between ensemble members.  Decoupling these spins may
be done by any number of pulse sequences which have been developed
over the past 40 years in solid-state NMR, the simplest of which
is the WAHUHA pulse sequence.  Such sequences can reduce dipolar
broadening by more than three orders of magnitude
\cite{haeberlin}.

\begin{figure}
\begin{center}
\epsfysize=1in \epsfbox{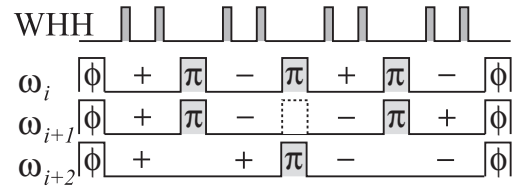}
\end{center}
\caption{An example RF pulse sequence.  The first line (WHH) shows
broadband decoupling of qubit copies; the narrow, shaded pulses
are appropriately phased $\pi/2$ pulses \protect \cite{haeberlin},
made sufficiently narrow (broadband) to affect all nuclei in the
crystal.  The subsequent three lines show selective pulses to
decouple qubits; these wider pulses are sufficiently long
(narrowband) to affect only individual planes.  Single spin
rotations, denoted by $\phi$, may be performed in between cycles.
Multi-bit gates such as controlled-NOT may be performed by
combining single spin rotations with selective recoupling.  For
example, qubits $i$ and $i+1$ may be recoupled by adding a $\pi$
pulse at the time noted by the dotted line \protect
\cite{leung99}. \label{pulsefig} }
\end{figure}

Fortunately, the different forms of $\hat{\cal H}_{ijmm}$ and
$\hat{\cal H}_{iimn}$ allow constant decoupling of copies without
adverse effect on the manipulations of couplings within individual
computers.  The qubit-qubit coupling of Eq. (\ref{ijcouple}) may
in general be rescaled and rotated by the broadband WAHUHA-type
sequence, but it may still be controlled for logic gates with
suitably phased $\pi$ pulses. The broadband WAHUHA-type pulse
sequence which decouples all copies is unaffected by the addition
of selective $\pi$-pulses, insofar as those $\pi$-pulses are
sufficiently short in comparison to the pulse spacing.  This
criterion is difficult to meet, since the $\pi$ pulses must last a
time of about $1/\Delta\omega_{i,i+1}$ to be frequency selective.
A simple example to illustrate the decoupling scheme is shown in
Fig. \ref{pulsefig}.  This example is limited by the adverse
effect of the finite $\pi$ pulse-length on the WAHUHA decoupling,
but longer and more sophisticated sequences can be designed with
standard techniques to compensate for this and other limiting
effects \cite{haeberlin}.

The only problematic dipolar couplings which remain after applying
the pulse sequences described above are those between qubits with
different resonant frequencies ($i\ne j$) and in different chains
($m\ne n$).  These couplings only arise when two planes of qubits
are selectively recoupled during a two-bit gate.   The needed time
duration for a logic operation may be minimized by maximizing the
desired qubit-qubit coupling -- this is accomplished by first
letting $a$ be the smallest nuclear distance in the crystal.
Second, we set $\phi=0$, hence putting the gradient parallel to
the applied field.  Third, only nearest-neighbor qubit couplings
are used ($j=i+1$), with longer distance couplings accomplished by
bit-swapping \cite{lloyd93}. After making these choices, we have
$\delta\omega\equiv\delta\omega_{i,i+1}=-\mu_0\gamma^2\hbar/2\pi
a^3$, and the target qubit of frequency $\omega_i$ in the $m$th
chain is coupled to control qubit copies at $\omega_{i+1}$ in the
other chains by the Hamiltonian
\begin{eqnarray}\nonumber
    \hat{\cal H}_{im}^*&=&-\hbar\delta\omega
    \sum_{n}\frac{\lambda_{mn}^2-2}{2(1+\lambda_{mn}^2)^{5/2}}
    {\hat I}_{im}^z {\hat I}_{i+1,n}^z \\
    &\equiv& -\hbar\delta\omega\sum_{n}b_{mn}{\hat I}_{im}^z {\hat
        I}_{i+1,n}^z.
\label{ijmncouple}\end{eqnarray} The decoherence caused by this
undesired Hamiltonian is minimized if the coupling constants are
sufficiently small to obey
\begin{equation}
\label{criterion}
    \frac{\sigma}{\delta\omega}=%\sqrt{1-\mu^2}
    \frac{1}{2}\sqrt{\sum_n b_{mn}^2}\ll 1,
\end{equation}
where $\sigma$ (the square root of the second moment
\cite{haeberlin}) is the effective linewidth of the qubit ensemble
during recoupling, which we demand to be much smaller than the
splitting $\delta\omega$. This inequality is best satisfied by a
crystal whose nuclei couple only in isolated chains.

A crystal which approximates this description is fluorapatite,
Ca$_5$F(PO$_4$)$_3$, whose $^{19}$F structure is shown in Fig.
\ref{fig}. In fluorapatite, we find $\sigma/\delta\omega\approx
1/58$, roughly six times smaller than in a simple cubic crystal
such as CaF$_2$.  The one-dimensional nuclear structure of
fluorapatite has been recognized in several NMR experiments
\cite{lugt64engelsberg73cho96}. Decoherence timescales in this
crystal are analagous to the well-known case of CaF$_2$.  The
$T_2$ time is limited by dipolar broadening; if dipolar couplings
are perfectly controlled, the timescale for internal decoherence
is pushed toward the spin-lattice relaxation time constant $T_1$,
which is limited by thermal fluctuations of paramagnetic
impurities (rare-earth substitutions, for example) and can easily
be several hours for reasonable crystal purities and temperatures
\cite{bloembergen49}.

The field gradient is accomplished by a micron-sized ferromagnetic
parallelopiped.  Calculations similar to those in Ref.
\cite{goldman2000} show that such a magnet, made of dysprosium and
placed in a 7~T external field as shown in Fig. \ref{fig}, can
produce a field gradient of 1.4~T/$\mu$m at a distance of 2.07
$\mu$m above the magnet, which leads to
$\Delta\omega\equiv\Delta\omega_{i,i+1}=2\pi\times 19.2$~kHz. A
1~$\mu$m by 10~$\mu$m fluorapatite crystal above the magnet
contains $N=10^7$ equivalent-frequency qubit copies in each
$xy$-plane. The field variation along the $y$-direction is
negligible since the magnet is much longer than the crystal. The
field variation along the $x$-axis is sufficiently small that all
of the equivalent-frequency nuclei lie within a bandwidth of
$\Delta\omega$.  We also note that inhomogeneous broadening is
constantly refocused in this scheme by the narrow-band $\pi$ pulse
sequences during both decoupling and selective recoupling.

The presence of a large magnetic field gradient provides a natural
means for performing MRFM on a magnetization $M^z$, since this
technique is sensitive to the gradient force given by $F=M^z\nabla
B^z$ \cite{sidles91%,rugar92
}. The crystal is mounted on a microfabricated cantilever which
oscillates in the $z$-direction, as in Fig. \ref{fig}.  The long
axis of the crystal-cantilever heterostructure and magnet are
aligned. The experiment is performed in high vacuum
($<10^{-5}$~torr) and at low temperatures. A coil is used to
generate the RF pulses for logic operations and decoupling
sequences; it also generates the continuous-wave radiation for
readout. An optical fiber-based displacement sensor is used to
monitor deflection of the cantilever using interferometry.
Sub-\AA{ngstrom} oscillations can be detected; larger oscillations
can be damped with active feedback which avoids additional
broadening while maintaining high sensitivity
\cite{durig97,cantileverfootnote}.

Readout is performed using cyclic adiabatic inversion
\cite{abragam}, which modulates the magnetization of a plane of
nuclei at a frequency near or on resonance with the cantilever.
The spins of resonant frequency $\omega_i$ are irradiated with the
RF field $B^{x} =
2B_{1}\cos\{\omega_{i}t-(\Omega/\omega_{m})\cos(\omega_{m}t)\}$,
where $\omega_{m}$ is the modulation frequency chosen to be near
the resonance of the cantilever, and $\Omega$ is the frequency
excursion, which should be much smaller than $\Delta\omega$
\cite{rugar94}. Simultaneous detection of signal from multiple
planes is possible if the different planes to be measured are
driven at distinct modulation frequencies $\omega_{m}$.

\begin{figure}
\begin{center}
\epsfysize=1.5in
%\epsfbox{figure3.eps}
\epsfbox{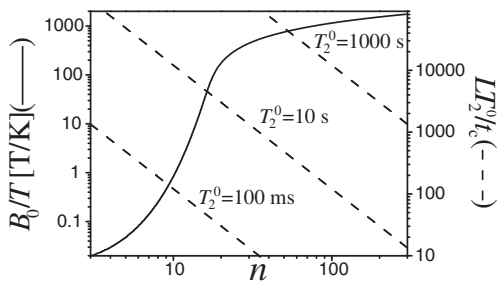}
\end{center} \caption{A
plot showing the scalability of the present scheme. The solid
curve, corresponding to the left axis, shows the magnetic field
(in T) divided by temperature (in K) needed in order for a number
of qubits $n$ to be measurable. The dashed lines, corresponding to
the right axis, plot the number of logic gates (decoherence time
$T_2$ divided by the pulse sequence cycle time $t_{\text{c}}$)
times the length of a ``block" of the decoupling sequence $L$
against $n$ for several values of $T_2^0.$ \label{scalingfig}}
\end{figure}

The force resolution for MRFM is limited by thermal fluctuations
of the cantilever \cite{gabrielson93}. Force resolutions of
$5.6\times10^{-18}$ N/$\sqrt{\rm Hz}$ have been reported for
single crystal cantilevers at 4 K \cite{yasamura2000}. To estimate
the force involved in a measurement following a quantum
computation, we consider the case in which initialization is
imperfect, so that the temperature $T$ is non-zero and an
effective pure state must be used \cite{nmrqc,knill98}. The usable
magnetization is estimated \cite{warren97} as
    $$M^z=
    \gamma\hbar\frac{N}{2^n}\frac{\sinh(n\hbar\gamma
    B_{0}/2k_{\text{B}}T)}{\cosh^{n}(\hbar\gamma
    B_{0}/2k_{\text{B}}T)}.$$
In the high temperature limit, this magnetization approximates to
$(\gamma^2\hbar^2 B_0/2k_{\text{B}}T) Nn2^{-n}$ showing the
exponential downscaling which plagues solution NMR. Even in this
limit, with the relatively small $N$ of $10^7$ and using the field
gradient described above in a 7~T field at 4~K, the force which
needs to be measured works out to about $10^{-15}
n2^{-n}$~newtons, allowing $n\sim 10$ qubits.  The density matrix
of such a system would be outside the neighborhood in which it has
been shown that all density matrices are separable
\cite{braunstein99}.  As seen in Fig. \ref{scalingfig}, the
situation is improved in larger fields and at lower temperatures,
where $M^z$ improves exponentially in $B_0/T.$  In the realistic
laboratory extreme $B_0=20$~T and $T=10$~mK, this design may scale
to $n\sim 300$ qubits; in this regime entanglement is shown to be
demonstrable \cite{braunstein99}.

A primary limitation to the long-term scalability of this scheme
is the number of logic gates which can fit into the decoherence
time when there is a large number of qubits. The
decoupling/recoupling pulse scheme based on Hadamard matrices has
a cycle time $t_{\text{c}}(n)$ of $Ln^2/\Delta\omega$, where the
parameter $L$ depends on the length of the homonuclear decoupling
subsequence and only weakly on the number of qubits $n$
\cite{leung99}.  Since single spin rotations must occur between
cycles, the clock speed is set by this cycle time.  The number of
logic gates we may perform is approximately $t_{\text{c}}(n)$
divided into the remaining decoherence time $T_2^0$.   This
timescale $T_2^0$ is due to residual homonuclear couplings,
cantilever thermal drift, and thermal relaxation resulting from
magnetic impurities; these contributions are limited by the
experimental performance of the homonuclear decoupling pulse
sequence, cantilever feedback stabilization, and crystal growth
technique, respectively.  Figure \ref{scalingfig} shows the number
of logic gates possible for several values of $T_2^0$.  The bottom
dashed trace ($T_2^0=100$~ms) corresponds to the case in which
homonuclear couplings are perfectly controlled but cantilever
drift is completely unsuppressed, while the middle ($T_2^0=10$~s)
and top ($T_2^0=1000$~s) traces assume suppression by 20 and 40 dB
with a negative feedback circuit.  The limited number of gates at
high numbers of qubits is a remaining problem which may be
addressed through error correcting codes \cite{shor95}, more
sophisticated pulse sequences, or the use of a more
one-dimensional system \cite{method}.

A remaining challenge in this scheme is the alignment of the field
gradient, cantilever, and crystal.  Such alignment is critical,
and although difficult, it is not impossible with modern
microfabrication technology and feedback control using
micro-electro-mechanical actuators \cite{miller97}.

This proposal features two important advantages over schemes using
solid-state NMR with impurity dopants \cite{kaneberman}. The
fabrication difficulty of artificially implanting controlled
arrays of spins is avoided in our proposal by using nuclei that
are naturally organized into a crystal structure. Also, the use of
an ensemble of 10$^7$ spins avoids the need for the daunting task
of reading and initializing single nuclear spins. Indeed, ensemble
measurement has given solution NMR a developmental head-start
against other existing proposals for quantum computation, and it
is hoped that the scheme presented here will carry NMR quantum
computation onward to the many-qubit regime.

This work was partially supported by NTT Basic Research
Laboratories.  T. D. L. was supported by the Fannie and John Hertz
Foundation.


\begin{references}
\bibitem{cantileverfootnote}
    A calculation using Redfield theory \cite{slichter} with the thermal noise statistics
    of a high-sensitivity cantilever \cite{gabrielson93}
    estimates the $T_2$ timescale due to the cantilever's thermal drift
    as $T_2= k \omega_{\text{c}}Qa^2/\Delta\omega^2 k_{\text{B}}
    T\approx 200$~ms, where $k\approx 10^{-3}$~N/m is the spring
    constant, $\omega_{\text c}/2\pi\approx$10~kHz is the resonant
    frequency, and $Q\approx 20,000$ is the quality factor of the
    cantilever.  Active feedback
    stabilization is expected to increase this timescale by as many as four
    orders of magnitude, bringing it close to the lengthy $T_1$
    timescale.
\bibitem{method}
    In a perfectly one-dimensional system with high $n$, the Hadamard
    decoupling scheme may be truncated to decouple sets of only
    $n_{\text{d}}<n$ nearest neighbors and then repeated periodically
    along a chain. The remaining couplings result in decoherence with
    a timescale $T_2^{\text{qq}}$ which scales as $n_{\text{d}}^3$,
    and since $t_{\text{c}}\sim n_{\text{d}}^2$, $T_2/t_{\text{c}}$
    may be made arbitrarily large with large enough $T_2^0$.  (In the
    real crystal fluorapatite, once $n_{\text{d}} > \lambda_{m,m+1} = 2.74,$ the
    %Note: character \lambda_{m,m+1} changed from undefined "c" in previous version
    contributions of the ensemble of copy-nuclei causes
    $T_2^{\text{qq}}\sim n_{\text{d}}^2$, leading to a constant
    $T_2^{\text{qq}}/t_{\text{c}}$ which is too small to be useful.)
\end{references}
\end{document}